\documentclass[12pt]{article}
\usepackage{young}
\usepackage{graphicx}
\usepackage{color}
\usepackage[latin1]{inputenc} 

\usepackage{amssymb}
\usepackage{amsfonts} 
\usepackage{amsthm} 
\usepackage{amsmath} 
\usepackage{hyperref}

\def\nn{\nonumber}

\def\ic{\mathrm{i}}

\def \bc {\begin{center}}
\def \ec {\end{center}}
\def \bi {\begin{itemize}}
\def \ei {\end{itemize}}

\def \ba {\begin{array}}
\def \ea {\end{array}}

\def \bea {\begin{eqnarray}}
\def \eea {\end{eqnarray}}

\def \be {\begin{equation}}
\def \ee {\end{equation}}

\def \lb {\left[}
\def \rb {\right]}

\def \um {\frac{1}{2}}
\def\tr {\mathrm{tr}}
\def\cD {{\cal D}}

\newcommand{\la}{\langle}
\newcommand{\ra}{\rangle}

\begin{document}

\begin{center}
{\Large {\bf Some properties of Grassmannian $U(4)/U(2)^2$ coherent states and an entropic conjecture}}
\end{center}
\bigskip

\centerline{{\sc Manuel Calixto}\footnote{Corresponding author: calixto@ugr.es}  
  and {\sc Emilio Pérez-Romero}}

\bigskip

\bc {\it  Departamento de Matemática Aplicada and Instituto Carlos I de F\'\i sica Te\'orica y Computacional,
Facultad de Ciencias, Universidad de Granada, Avenida Fuentenueva s/n,  18071 Granada, Spain}
\ec

\bigskip
\begin{center}
{\bf Abstract}
\end{center}
\small
\begin{list}{}{\setlength{\leftmargin}{3pc}\setlength{\rightmargin}{3pc}}
\item

We analyze mathematical and physical properties of a previously introduced [J. Phys. A47, 115302 (2014)] family of $U(4)$ coherent states (CS). 
They constitute a matrix version of standard spin $U(2)$ CS when we add an extra (pseudospin) dichotomous degree of freedom: 
layer, sublattice, two-well, nucleon, etc. Applications to bilayer quantum Hall systems at fractions 
of filling factor $\nu=2$ are discussed, where Haldane's sphere picture is generalized to a Grassmannian picture. 
We also extend Wehrl's definition of entropy from Glauber to Grassmannian CS and state a conjecture on the 
entropy lower bound.

\end{list}
\normalsize 




\section{Introduction}

Spin-$s$, Bloch, $SU(2)$ or atomic coherent states  (CS) were introduced by Radcliffe \cite{Radcliffe71}, Gilmore \cite{Gilmore1,Gilmore11} 
and Perelomov \cite{Perelomov}. They, together with standard (Glauber) CS \cite{Glauber}, accurately describe the physical properties of many macroscopic quantum systems like in: quantum optics, 
Bose-Einstein condensates (BEC), two-level systems, superconductors, superfluids, quantum Hall effects, etc; see e.g. \cite{Klauder,Perelomovbook,Dodonov,Gazeaubook,JPAissue,Antoine} for a selected panorama 
of applications (not only in quantum mechanics, but also in engineering). In particular, the ground state of many physical systems undergoing a quantum phase transition (QPT) is well described by a CS. 
Actually, it was Gilmore who introduced an algorithm \cite{Gilmore2}, which makes use of CS as variational
states to approximate the ground state energy, to study the {\it classical, thermodynamic or meanfield limit} of some algebraic quantum models. This algorithm has proved to be specially
suitable to analyze the phase diagram of Hamiltonian models undergoing a QPT like: the Dicke model 
of atom-field interactions  \cite{Dicke,CastaPRA,DickeEPL,DickePRA}, Bose-Einstein condensates \cite{AnalCasta},  the Lipkin-Meshkov-Glick model \cite{LMG,CastaPRB,LipkinPS,LipkinEPL}, vibron model for molecules 
\cite{vibroncurro,VibronHusimiPRA,EntangVibronPRA}, etc. In some quantum phases, the ground state is in fact a (parity) symmetry adapted CS or ``Schr\"odinger cat'' \cite{Dodonovcat,Mankocat}.

The generalization from $U(2)$ to $U(N)$ CS is quite straightforward for the fully symmetric (canonical) case, related to the complex projective $\mathbb{C}P^{N-1}=U(N)/[U(N-1)\times U(1)]$ quotient space. 
However, general Grassmannian cosets  $\mathbb{G}^N_k=U(N)/[U(N-k)\times U(k)]$, related to other less symmetric Young tableau arrangements, are a bit more involved and perhaps less known or explicitly worked out. 
Actually, we are interested in the noncanonical chain of subgroups
\[U(4)\supset U(2)\times U(2)\supset U(1)\times U(1).\]
The corresponding matrix representation is useful, for instance, when adding a pseudospin (layer, two-well, nucleon, sublattice, etc) to the spin degree of freedom. For example, in 
bilayer spin (namely, quantum Hall) systems, pseudospin (pspin for short) is introduced in order to emphasize
the spin $SU(2)$ symmetry in the, let us say,  bottom (pspin $-1/2$) and top (pspin $1/2$) layers \cite{EzawaBook,GrassCSBLQH,CalixtoJPCM14}; 
here  pspin rotates when particles are transfered from one layer to the other. Also, as noted long time ago by Heisenberg and Wigner \cite{WignerU4}, 
$U(4)$ (which assigns isotopic spin $-1/2$ to protons and $1/2$ to neutrons) is an approximate symmetry of nuclear forces when neglecting 
electromagnetic forces. Moreover, pspin in graphene makes reference to the two triangular sublattices  of the two-dimensional honeycomb structure.

We shall make use of the bilayer quantum Hall (BLQH) picture all along this paper to provide a physical interpretation of some of the mathematical structures that will arise. 
In fact, spin-$s$ CS already appear in Haldane's sphere picture \cite{Haldane} for the (monolayer) fractional quantum Hall effect, in which $2s$ makes reference to total magnetic flux through the surface 
in units of the Dirac magnetic flux quantum $\Phi=2\pi\hbar/e$; the spin $s$ is also related to the ``monopole strength'' in the sphere $\mathbb S^2$. 
This image has to do with the ``composite fermion'' picture of Jain \cite{Jainbook}, which explains the existence of fractional 
values for the filling factor $\nu$ (number of electrons per magnetic flux quantum penetrating the sample). The generalization of Haldane's sphere $\mathbb S^2=U(2)/U(1)^2$ from monolayer to bilayer quantum Hall systems is the 
Grassmannian $\mathbb{G}_2^4=U(4)/U(2)^2$ (we simply write $\mathbb{G}_2$, since no confusion will arise) for fractional values of the filling factor $\nu=2/\lambda$, with the isospin $\lambda$ related to the number of magnetic flux 
quanta piercing each of the two electrons (also the monopole strength in $\mathbb G_2$); see \cite{GrassCSBLQH,CalixtoJPCM14} for more information. 

The organization of the paper is as follows. Firstly in Section \ref{Bloch} we briefly review the simpler case of $U(2)$ CS, which allows us to better understand the more involved $U(4)$ case. 
In Sec. \ref{GrassCSec} we introduce different bases of $U(4)$ operators, the carrier Hilbert space of the unirrep associated to the phase space $U(4)/U(2)^2$ and the corresponding set of 
CS. We show that Grassmannian CS can be seen as a matrix generalization of Bloch CS with a generalized binomial distribution (Sec. \ref{Binomsec}). In Sec. \ref{symbolsec} 
we compute operator symbols and relate them to order parameters in the BLQH jargon, 
studying the so called spin, pspin and canted quantum phases \cite{EzawaBook}. Using the CS (Bargmann) representation (see \cite{Vourdas} for a review), in Sec. \ref{Husisec} we also compute the Husimi function $Q_\psi$ of any state $\psi$ and extend 
Wehrl's definition of entropy \cite{Wehrl79} from Glauber to Grassmannian CS. We conjecture a lower bound for this extension of Wehrl's entropy (and an upper bound for the Husimi's second moment) which is attained for any 
Grassmannian CS. Finally, Section \ref{comments} is left for conclusions.

\section{Bloch $U(2)/U(1)^2$ spin-$s$ coherent  states: a brief}\label{Bloch}

Let us consider the carrier space ${\cal H}_s$ of a $(2s+1)$-dimensional irrep of $SU(2)$ spanned by the orthonormal basis vectors 
$\{|s,k\rangle, k=-s,\dots,s\}$, which are eigenstates of the angular momentum third component $J_3$ with eigenvalue $k$. From now on we shall remove the spin label 
$s$ from the basis vectors and simply denote $|s,k\rangle=|k\rangle$. 
Spin-$s$ CS  can be obtained as an exponential action of angular momentum ladder operators $J_\pm=J_1\pm\ic J_2$ 
on (namely) the lowest-weight state $|k\ra=|-s\ra$ as
\be
|z\rangle=\frac{e^{\bar z {J}_+}|-s\rangle}{(1+|z|^2)^{s}}=
\frac{\sum_{k=-s}^s\varphi_k(\bar z)|k\rangle}{(1+|z|^2)^{s}},\label{su2cs}
\ee
with  $\varphi_k(z)=\binom{2s}{s+k}^{1/2}z^{s+k}$ a set of monomials in the complex variable $z$ verifying the closure relation (the Bergmann kernel $K_{2s}$)
\be
\sum_{k=-s}^s\varphi_k(z')\varphi_k(\bar z)=(1+z'\bar z)^{2s}\equiv K_{2s}(z',\bar z).\label{BK}
\ee
The complex label $z$ is also usually parametrized as
$z=\tan(\theta/2)e^{\ic\phi}$, related to the stereographic projection of a
point $(\theta,\phi)$ (polar and azimuthal angles) of the Riemann sphere $\mathbb S^2=U(2)/U(1)^2$ onto the complex plane. The probability 
of measuring the angular momentum third component $k$ in the CS $|z\rangle$ follows a binomial distribution
\begin{equation}
 |\langle k|z\rangle|^2=\frac{|\varphi_k(z)|^2}{(1+|z|^2)^{2s}}=\binom{2s}{s+k}\left(\cos^2\frac{\theta}{2}\right)^{s+k}\left(\sin^2\frac{\theta}{2}\right)^{s-k},\label{binomdist}
\end{equation}
with parameter $\cos^2({\theta}/{2})$. This binomial structure is also apparent from 
another familiar Fock-space representation [equivalent to \eqref{su2cs}] of spin-$s$
CS as a two-mode, $a$ and $b$, Bose-Einstein condensate (BEC) of $N=2s$ particles, given by the expression 
($|0_F\rangle$ denotes the Fock vacuum)
\be
|z\rangle=\frac{1}{\sqrt{(2s)!}}\left(\frac{b^\dag+\bar z a^\dag}{\sqrt{1+|z|^2}}\right)^{2s}|0_F\ra.\label{su2BE}
\ee
In bilayer quantum Hall (BLQH) systems --spin-frozen case-- and in Bose-Einstein condensates in a double well potential, the operators 
$a^\dag$ and $b^\dag$ create (flux) quanta in layers/wells $a$ and $b$, respectively \cite{CalixtoJPCM14,AnalCasta}.  In these physical 
contexts, the polar angle $\theta$ is related to the population imbalance between wells/layers (see \cite{CalixtoJPCM14} and later on this section), represented by the operator 
$J_3=\um(a^\dag a-b^\dag b)$ in the usual  Jordan-Schwinger boson realization  for spin $J_\mu=\um\begin{pmatrix} a^\dag & b^\dag \end{pmatrix}\sigma_\mu\begin{pmatrix} a & b \end{pmatrix}^t$,  
where $\sigma_\mu$ denote the usual three Pauli matrices 
($\mu=1,2,3$) plus the $2\times 2$ identity matrix $\sigma_0$, and $2J_0=a^\dag a+b^\dag b$ is the total number of particles. 
The azimuthal angle $\phi$ is the relative phase between the two spatially separated wells/layers, an important magnitude to codify qubits in these systems.

Spin-$s$ CS \eqref{su2cs} are normalized,  but not orthogonal in general, as can be seen from the CS overlap \be\langle z'|z\rangle=\frac{K_{2s}(z',\bar z)}{K_s(z',\bar z')K_s(z,\bar z)},\label{su2overlap}\ee
written in terms of the Bergmann kernel \eqref{BK}. Spin-$s$ CS constitute an overcomplete 
set and fulfills  the resolution of unity $1=\int_{{\mathbb S^2}}|z\rangle\langle z|d\mu(z,\bar z)$,
with measure $d\mu(z,\bar z)=\frac{2s+1}{\pi}\frac{d^2z}{(1+|z|^2)^2}=\frac{2s+1}{4\pi}\sin\theta d\theta d\phi$ (the solid angle element).

We shall make extensive use of the CS (Fock-Bargmann) representation $\Psi(z)$ (the ``symbol'') of any state $|\psi\rangle\in {\cal H}_s$, defined as 
\begin{equation}
 \Psi(z)\equiv K_s(z,\bar z)\langle z|\psi\rangle.\label{FBrep}
\end{equation}
For example, from \eqref{su2cs} and \eqref{FBrep}, the 
basis states $|\psi\rangle=|k\rangle$ are represented by the monomials $\varphi_k(z)$ of degree $s+k$, whereas general states 
$|\psi\rangle=\sum_k c_k|k\rangle$ are represented by polynomials $\Psi(z)=\sum_k {c}_k\varphi_k(z)$ of degree $\leq 2s$ in the variable $z$. 
The relation \eqref{FBrep} makes the space  ${\cal H}_s(\mathbb S^2)$ of polynomials of degree $2s$ in the variable $z$ a \emph{reproducing kernel Hilbert space} 
[that is, a Hilbert space of functions $\Psi$ in which pointwise evaluation $\Psi(z)$ is a continuous linear functional] with measure 
$d\mu_s(z,\bar z)\equiv d\mu(z,\bar z)/K_{2s}(z,\bar z)$. Inside this 
CS picture, angular momentum operators $J_i$ are represented by differential operators $\mathcal{J}_i$, namely 
\begin{equation}
 \mathcal{J}_3=z\frac{d}{dz}-s,\;  \mathcal{J}_+=-z^2\frac{d}{dz}+2sz,\;  \mathcal{J}_-=\frac{d}{dz},
\end{equation}
so that the following identity holds: $\mathcal{J}_i\Psi(z)= K_s(z,\bar z)\langle z|J_i|\psi\rangle$. This relation is useful for technical 
calculations; for example, taking $|\psi\rangle=|z'\rangle$, we can easily compute CS matrix elements as
\begin{equation}
 \langle z|J_i|z'\rangle=\frac{\mathcal{J}_iK_{2s}(z,\bar z')}{K_{s}(z,\bar z)K_{s}(z',\bar z')}.
\end{equation}
In particular, the diagonal elements $\langle{J}_i\rangle\equiv\langle z|J_i|z\rangle$ (also called ``lower'' or ``contravariant'' symbols of $J_i$) 
are simply $\langle{J}_i\rangle=K_{2s}^{-1}(z,\bar z)\mathcal{J}_iK_{2s}(z,\bar z)$. In the same way, CS expectation values of operator products can be computed as 
$\langle J_jJ_i\rangle=K_{2s}^{-1}(z,\bar z)\mathcal{J}_j(\mathcal{J}_iK_{2s}(z,\bar z))$. For example, in the context of BEC/BLQH systems 
commented after \eqref{su2BE}, the quantities 
\begin{equation}
 \la{J}_3\ra=s\frac{|z|^2-1}{|z|^2+1}=-s\cos\theta,\;  \la{J}_3^2\ra=\la{J}_3\ra^2+\frac{2s|z|^2}{(|z|^2+1)^2}=s^2\cos^2\theta+
 \frac{s}{2}\sin^2\theta,
\end{equation}
are related to population imbalance $\la J_3\ra$ and its fluctuations $\sqrt{\la J_3^2\ra-\la J_3\ra^2}$ between wells/layers.
As commented at the introduction, these CS expectation values are important quantities to compute energy surfaces and to study 
the  classical, mean-field or thermodynamical limit of spin systems undergoing a quantum phase transition (QPT). Actually, spin CS  saturate the Heisenberg 
uncertainty relation $\la J_1^2\ra\la J_2^2\ra\geq\frac{1}{4}\la J_3\ra^2$ and are therefore ``close to classical''.  Semiclassical properties 
of quantum spin states given by a density matrix $\rho$ are also described by the CS expectation value $Q(z,\bar z)=\la z|\rho|z\ra$, also called 
Husimi quasiprobability distribution, lower symbol or 
$Q$-function of $\rho$. For pure states, $\rho=|\psi\ra\la\psi|$, we have $Q(z,\bar z)=|\la z|\psi\ra|^2=|\Psi(z)|^2/K_{2s}(z,\bar z)$. If $\la \psi|\psi\ra=1$, then 
$Q$ is normalized according to $\int_{\mathbb{S}^2}Q(z,\bar z)d\mu(z,\bar z)=1$. 
Wehrl's entropy $W=\int_{\mathbb{S}^2} Q(z,\bar z)\ln Q(z,\bar z)d\mu(z,\bar z)$ measures the area 
occupied by a state $\psi$ in phase space, which is proved to be minimal, $W_{\mathrm{min}}=1-1/(2s+1)$,   
when $\psi$ is itself a CS $|\psi\rangle=|z'\ra$ (see \cite{LiebAM} for a recent proof, which was conjectured in 1978 by Lieb \cite{LiebCMP}). 
Equivalently, the second moment (``inverse participation ratio'') $M=\int_{\mathbb{S}^2}Q^2(z,\bar z)d\mu(z,\bar z)$ is maximum, $M_{\mathrm{max}}=1/2+1/(2+8s)$, when $\psi$ is a 
CS. In the large spin limit \cite{Holstein,Radcliffe71} we have  $W_{\mathrm{min}}\to 1$ and $M_{\mathrm{max}}\to 1/2$, thus recovering the extremal values attained by ordinary Heisenberg-Weyl (harmonic oscillator) coherent states, 
as proved by Wehrl \cite{Wehrl79}. 
We shall also conjecture extremal values of $W$ and $M$ for $U(4)/U(2)^2$ CS in the next section. 

To conclude this section, let us remark that all these information measures (R\'enyi-Wehrl entropies) have been of important use to characterize QPTs of 
several molecular and optical models \cite{DickeHusimiPRA,VibronHusimiPRA,EntangVibronJPA,EntangVibronPRA,LipkinPS,DickePS} and topological insulators \cite{calixto15,romera15,calixto15JSTAT}.

There are many more interesting mathematical properties and physical applications of $U(2)/U(1)^2$ CS, but we just want to give here a brief account enough to introduce and compare 
with the more involved $U(4)/U(2)^2$ CS in the next section.

\section{Grassmannian $U(4)/U(2)^2$ isospin-$\lambda$ coherent states}\label{GrassCSec}


We shall add an extra pseudospin (pspin) degree of freedom to spin, which will lead to some matrix generalizations of the 
standard spin-$s$ CS expressions of the previous section. Looking for physical applications, pspin may refer to a ``layer, well, sublattice, nucleon, etc'' 
extra degree of freedom that comes into play.

\begin{figure}
\begin{center}
\includegraphics[width=6cm]{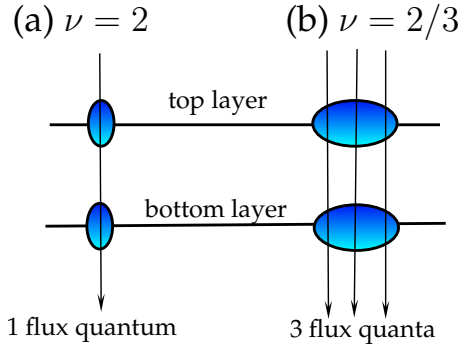}
\end{center}
\caption{Composite particles (bosons or fermions), seen as ``fat electrons'' bound to $\lambda$ (even or odd) magnetic flux lines, explain fractional quantum Hall effect at filling 
factors $\nu=n/\lambda$. We represent $n=2$ electrons in a bilayer system pierced by $\lambda=1$ (a) and $\lambda=3$ (b) magnetic flux lines, leading to $\nu=2/1$ and $\nu=2/3$, respectively.}
\label{fig0}
\end{figure}

To emphasize the spin-pspin (combined into ``isospin'', to use the same nomenclature as in 
\cite{EzawaBook}) structure, we denote the $U(4)$ generators in the fundamental representation by the sixteen $4\times 4$ matrices 
$\tau_{\mu\nu}\equiv\sigma_\mu^\mathrm{pspin}\otimes\sigma_\nu^\mathrm{spin}, \, \mu,\nu=0,1,2,3$ (we shall use Greek letters $\mu,\nu$ for four-indices and latin  
letters $i,k,l=1,2,3$ for the ``spatial'' part). General $U(4)$ unirreps arise 
in the Clebsch-Gordan decomposition of a tensor product of $N$ four-dimensional (fundamental, elementary) representations of
$U(4)$, for example:
\be  \overbrace{\begin{Young}  \cr \end{Young}  \otimes\dots\otimes \begin{Young}  \cr \end{Young}}^{N=2\lambda}=
 \overbrace{\begin{Young}  & ... & \cr  \end{Young}}^{2\lambda}\oplus
   \overbrace{\begin{Young}  & ... & \cr & ...  & \cr \end{Young}}^{\lambda}\oplus\dots\quad  \mathrm{or}\quad
   \overbrace{[1]\otimes\dots\otimes [1]}^{N=2\lambda}=[N]\oplus[\lambda,\lambda]\oplus\dots\label{CGdecomp}
\ee
The best known case is the fully symmetric unirrep [first term in the direct sum decomposition \eqref{CGdecomp}] corresponding to the Young Tableau of shape $[N]$ (one row of $N$ boxes) with 
dimension $(N+1)(N+2)(N+3)/6$. Here we are interested in those  unirreps corresponding to the Young Tableau of shape 
$[\lambda,\lambda]$ with two rows of $\lambda=N/2$ (for even $N$) boxes each [second term in the direct sum decomposition \eqref{CGdecomp}]. 
The dimension of the tableau  $[\lambda,\lambda]$ is  $d_\lambda=\frac{1}{12}(\lambda+1)(\lambda+2)^2(\lambda+3)$. 
For example, for $\lambda=1$ ($N=2$) we have $[1]\otimes [1]=[2]\oplus[1,1]$ (that is $4\times 4=10+6$)
so that $[1,1]$ corresponds to the fully antisymmetric irrep with dimension $d_1=6$. 
In Refs. \cite{GrassCSBLQH,CalixtoJPCM14} we have provided  a ``composite particle'' picture (a term imported from the
quantum Hall effect jargon \cite{Jainbook}) to physically interpret the $[\lambda,\lambda]$ configurations as all possible ways of distributing $2\lambda$ flux quanta 
among two identical fermions with spin 1/2 occupying two layers $a$ (top) and $b$ (bottom); see Figure \ref{fig0} for an illustration. Here we reproduce the argument, which is the following. 
We have two electrons attached to $\lambda$ flux quanta each.
The first electron can occupy any of the four isospin (layer, spin) states:  $|b\!\uparrow\rangle, |b\!\downarrow\rangle,
|a\!\uparrow\rangle$ and $|a\!\downarrow\rangle$, at one Landau site of the lowest Landau level. Therefore, there are $\tbinom{4+\lambda-1}{\lambda}$ ways of
distributing $\lambda$ quanta among these four states. Due to the Pauli exclusion principle, there are only three states
left for the second electron and $\tbinom{3+\lambda-1}{\lambda}$ ways of
distributing $\lambda$ quanta among these three states. However, some of the previous configurations must be identified
since both electrons are indistinguishable and $\lambda$ pairs of quanta adopt $\tbinom{2+\lambda-1}{\lambda}$
equivalent configurations. In total, there are
\begin{equation}
 {\binom{\lambda+3}{\lambda}\binom{\lambda+2}{\lambda}}/{\binom{\lambda+1}{\lambda}}=\frac{1}{12}(\lambda+3)(\lambda+2)^2(\lambda+1)=d_\lambda
\end{equation}
ways to distribute $2\lambda$ flux quanta among two identical electrons in four states, which turns out to coincide with
the dimension $d_\lambda$ of the tableau  $[\lambda,\lambda]$. In the BLQH jargon, this case corresponds to filling factor $\nu=2/\lambda$. The case $\nu=1/N$ (one electron) is related to the fully symmetric 
representation $[N]$ with dimension $(N+1)(N+2)(N+3)/6$, which coincides with the number of ways $\tbinom{N+3}{N}$ of distributing $N$ flux quanta among four isospin states.

The corresponding carrier Hilbert space ${\cal H}_\lambda$ of the $d_\lambda$-dimensional 
unirrep of $U(4)$ is spanned by the set of orthonormal basis vectors $\{|{}{}_{q_a,q_b}^{j,m}\ra, {2j+m\leq\lambda}, |q_{a,b}|\leq j\}$ fulfilling the resolution of the identity \cite{GrassCSBLQH} 
\be
 1=\sum^{\lambda}_{m=0}\sum_{j=0;\um}^{(\lambda-m)/2}\sum^{j}_{q_a,q_b=-j}
|{}{}_{q_a,q_b}^{j,m}\ra \la{}{}_{q_a,q_b}^{j,m}|,\label{basisvec}
\ee
where $\sum_{j=0;\um}$  means sum on $j=0,\um,1,\frac{3}{2},\dots$ (the ``angular momentum''). 
These are the $U(4)$ isospin-$\lambda$ analogue of the $U(2)$ spin-$s$  orthonormal
basis vectors $|k\rangle$ of the previous section (we are omitting the labels $s$ and $\lambda$ from the basis vectors $|k\rangle$ and $|{}{}_{q_a,q_b}^{j,m}\ra$, respectively,
for the sake of brevity). The sixteen $U(4)$ generators $\tau_{\mu\nu}$ are represented in ${\cal H}_\lambda$ by sixteen operators $T_{\mu\nu}$ whose matrix elements have been calculated in 
Ref.  \cite{GrassCSBLQH}. It is convenient to introduce the $U(2)^2\subset U(4)$ (antisymmetric Lorentz-like)  generators $M_{\mu\nu}$ by $M_{ik} =-\frac{\ic}{2}\epsilon_{ikl}T_{0l}$ and $M_{0i}=-\um T_{3i}$, $i,k,l=1,2,3$ and  
the isospin ladder operators by $T_{\pm\mu}=(T_{1\mu}\pm \ic T_{2\mu})/2$. We shall use the Einstein 
summation convention with Minkowski metric $\eta_{\mu\nu}=\mathrm{diag}(1,-1,-1,-1)$ (unless otherwise stated) and denote, for example, $\check T_{\pm\mu}\equiv T_{\pm}^\mu=
\eta^{\mu\nu}T_{\pm\nu}$. In the BLQH literature (see e.g.
\cite{EzawaBook}) it is customary to denote the total spin ${S}_i={T}_{0i}/2=\ic\epsilon_{ikl}M_{kl}$ and
pspin ${P}_i={T}_{i0}/2$, together with the remaining 9 isospin  ${R}_{ki}={T}_{ik}/2$ operators.
Angular momentum operators $S_{\ell i}$ of layers $\ell=a,b$ (resp. pspin $\ell=\mp$) are then given by 
\begin{equation}
 S_{\ell\, i}=\frac{1}{4}(\mp{T}_{0i}-{T}_{3i})=\um(M_{0i}\mp\frac{\ic}{2}\epsilon_{ikl}M_{kl})=\um (\mp S_i-R_{i3}),\label{spinab}
\end{equation}
so that $\vec{S}=\vec{S}_b-\vec{S}_a$ and  $M^2=M_{\mu\nu}M^{\mu\nu}=-4(\vec{S}_a^2+\vec{S}_b^2)=-2(\vec{S}^2+\vec{R}_3^2)$ with $\vec{R}_i=(R_{1i},R_{2i},R_{3i})$,  
$\vec{S}=(S_1,S_2,S_3)$ and $\vec{P}=(P_1,P_2,P_3)$. The linear and quadratic 
$U(4)$ Casimir operators are $C_1=T_{00}$ and 
\bea
C_2&=&\frac{1}{4}\delta^{\mu\nu}\delta^{\alpha\beta}T_{\mu\alpha}T_{\nu\beta}-\frac{1}{4}T^2_{00}
 =\frac{1}{4}T_{30}^2-\um M^2+\um(\check{T}_{-\mu}
 {T}_+^\mu+\check{T}_{+\mu} T_{-}^\mu)\nn\\
 &=& \vec{S}^2+\vec{P}^2+\mathbf{R}^2,\label{Casimir}
 \eea
respectively, with $\mathbf{R}^2=\vec{R}_1^2+\vec{R}_2^2+\vec{R}_3^2$ and $\delta^{\mu\nu}$ the Kronecker delta. 

The basis vectors $|{}{}_{q_a,q_b}^{j,m}\ra$ are eigenstates of $P_{3}, S_{\ell 3}$ and $M^2$ with eigenvalues
\begin{equation}
 P_{3}|{}{}_{q_a,q_b}^{j,m}\ra=(2j+m-\lambda)|{}{}_{q_a,q_b}^{j,m}\ra,\;  S_{\ell 3}|{}{}_{q_a,q_b}^{j,m}\ra=q_{\ell}|{}{}_{q_a,q_b}^{j,m}\ra,\;  M^2|{}{}_{q_a,q_b}^{j,m}\ra=-8j(j+1)|{}{}_{q_a,q_b}^{j,m}\ra. 
\end{equation}
We also have
\be
C_1|{}{}_{q_a,q_b}^{j,m}\ra=2\lambda |{}{}_{q_a,q_b}^{j,m}\ra, \; C_2|{}{}_{q_a,q_b}^{j,m}\ra=\lambda(\lambda+4) |{}{}_{q_a,q_b}^{j,m}\ra.
\ee
The generalization of spin-$s$ CS $|z\ra$ in \eqref{su2cs} to isospin-$\lambda$ CS $|Z\ra$ can be formally accomplished by, roughly speaking, substituting some scalars by $2\times 2$ matrices. 
In particular, the complex scalar $z\in\mathbb{S}^2=U(2)/U(1)^2$ is replaced by a $2\times 2$ complex matrix $Z\in\mathbb{G}_2=U(4)/U(2)^2$, that is, $Z=z^\mu\sigma_\mu$, with four
complex coordinates  $z^\mu=\tr(Z\sigma_\mu)/2\in\mathbb{C}$.  Sometimes we will also adopt the following decomposition for
a matrix $Z\in\mathbb{G}_2$
\be Z=V_a \begin{pmatrix}\xi_+  &0\\ 0& \xi_- \end{pmatrix} V_b^\dag,\; \xi_\pm=\tan\frac{\vartheta_\pm}{2}e^{\ic\beta_\pm},\;
V_\ell=\begin{pmatrix}\cos\frac{\theta_\ell}{2}&-\sin\frac{\theta_\ell}{2}e^{\ic\phi_\ell}\\ \sin\frac{\theta_\ell}{2}e^{-\ic\phi_\ell} & \cos\frac{\theta_\ell}{2} \end{pmatrix},\, \ell=a,b, \label{paramang}
\ee
in terms of the eight angles $\theta_{a,b},\vartheta_\pm\in[0,\pi)$ and $\phi_{a,b},\beta_{\pm}\in[0,2\pi)$, where $V_{a,b}$ represent rotations in layers $\ell=a,b$ (note their ``conjugated'' character).

Isospin-$\lambda$ CS can be obtained as an exponential action of ladder operators $T_\pm=T_{\pm\mu}\sigma^\mu$ on   
(namely) the lowest-weight state $|{}{}_{q_a=0,q_b=0}^{j=0,m=0}\ra$  and  can be expanded in terms of
the orthonormal basis vectors $|{}{}_{q_a,q_b}^{j,m}\ra$  as
\be
|Z\ra=\frac{e^{\um \tr(Z^\dag{T}_{+})}|{}{}_{0,0}^{0,0}\ra}{\det(\sigma_0+ZZ^\dag)^{\lambda/2}}=\frac{\sum^{\lambda}_{m=0}\sum_{j=0;\um}^{(\lambda-m)/2}
\sum^{j}_{q_a,q_b=-j}\varphi_{q_a,q_b}^{j,m}(\bar Z)
|{}{}_{q_a,q_b}^{j,m}\ra}{\det(\sigma_0+ZZ^\dag)^{\lambda/2}},\label{u4cs}
\ee
with 
\be
\varphi_{q_a,q_b}^{j,m}(Z)=\sqrt{\frac{2j+1}{\lambda+1}\binom{\lambda+1}{2j+m+1}\binom{\lambda+1}{m}} \det(Z)^{m}\cD^{j}_{q_a,q_b}(Z),\label{basisfunc}\ee
a set of homogeneous polynomials of degree $2j+2m$ in four complex variables $z^\mu$, generalizing the $U(2)$ monomials $\varphi_k(z)$ in \eqref{su2cs}. 
They are written in terms of the usual Wigner $\cD$-matrix \cite{Louck3}
\be
 \cD^{j}_{q_a,q_b}(Z)=\sqrt{\frac{(j+q_a)!(j-q_a)!}{(j+q_b)!(j-q_b)!}}
 \sum_{k=\max(0,q_a+q_b)}^{\min(j+q_a,j+q_b)}\tbinom{j+q_b}{k}\tbinom{j-q_b}{k-q_a-q_b}   z_{11}^k
z_{12}^{j+q_a-k}z_{21}^{j+q_b-k}z_{22}^{k-q_a-q_b},\nn \ee
for a general $2\times 2$ complex matrix $Z$ with entries $z_{uv}$ and angular momentum $j$. Homogeneous polynomials \eqref{basisfunc} verify the
closure relation (the Bergmann kernel $K_\lambda$)
\be\sum^{\lambda}_{m=0}\!\!\sum_{j=0;\um}^{(\lambda-m)/2}\!\!\sum^{j}_{q_a,q_b=-j}\!\!
{\varphi_{q_a,q_b}^{j,m}({Z'})}\varphi_{q_a,q_b}^{j,m}(\bar Z)={\det(\sigma_0+Z'
Z^\dag)^\lambda}\equiv K_\lambda(Z',Z^\dag)\label{kernel}\ee
from which we see that isospin-$\lambda$ CS are normalized but not orthogonal, with CS overlap 
\be
\la Z'|Z\ra=\frac{K_\lambda(Z',Z^\dag)}{K_{\lambda/2}(Z',Z'^\dag)K_{\lambda/2}(Z,Z^\dag)}.\label{u4csov}
\ee
Using  orthogonality properties of the homogeneous polynomials
$\varphi_{q_a,q_b}^{j,m}(Z)$, a resolution of unity for isospin-$\lambda$ CS has been proved in \cite{GrassCSBLQH}, namely
$1=\int_{\mathbb G_2} |Z\ra\la Z|d\mu(Z,Z^\dag)$, with integration measure 
\be
 d\mu(Z,Z^\dag)=\frac{12d_\lambda}{\pi^4}\frac{\prod_{\mu=0}^3 d^2z^\mu}{\det(\sigma_0+Z^\dag Z)^{4}}=\frac{3d_\lambda}{2^9\pi^4}
 (\cos\vartheta_+-\cos\vartheta_-)^2d\varOmega_+ d\varOmega_- d\Omega_a d\Omega_b,\label{measureG2}
 \ee 
where $d\varOmega_\pm=\sin\vartheta_\pm d\vartheta_\pm d\beta_\pm$ and $d\Omega_{\ell}=\sin\theta_{\ell} d\theta_{\ell} d\phi_\ell$ ($\ell=a,b$) are  solid angle elements. 
This is the $\mathbb{G}_2$ generalization of the $\mathbb{S}^2$ integration measure given after eq. \eqref{su2overlap}.

\subsection{A matrix generalization of the binomial distribution}\label{Binomsec}

The probability of finding $|{}{}_{q_a,q_b}^{j,m}\ra$ in a CS $|Z\ra$ follows the multivariate distribution function
\bea
Q_{q_a,q_b}^{j,m}(Z,Z^\dag)&\equiv&|\la Z|{}{}_{q_a,q_b}^{j,m}\ra|^2=\frac{|\varphi_{q_a,q_b}^{j,m}(Z)|^2}{\det(\sigma_0+ZZ^\dag)^{\lambda}}\nn\\ &=&
{\frac{2j+1}{\lambda+1}\binom{\lambda+1}{2j+m+1}\binom{\lambda+1}{m}} \frac{\det(ZZ^\dag)^{m}}{\det(\sigma_0+ZZ^\dag)^{\lambda}}|\cD^{j}_{q_a,q_b}(Z)|^2.\label{husibasis}
\eea
This is the Husimi function of a basis state $|{}{}_{q_a,q_b}^{j,m}\ra$ (see later on Sec. \ref{Husisec}). Let us study several particular cases. For example, 
using the parametrization \eqref{paramang}, the Husimi function of the lowest-weight state $|{}{}_{q_a,q_b}^{j,m}\ra=|{}{}_{0,0}^{0,0}\ra$ is given by the simple expression
 \be Q_{0,0}^{0,0}(Z,Z^\dag)=\left(\cos^{2}\frac{\vartheta_+}{2}\cos^{2}\frac{\vartheta_-}{2}\right)^\lambda,\label{Qlowest}\ee
 whereas for $|{}{}_{q_a,q_b}^{j,m}\ra=|{}{}_{0,0}^{0,m}\ra$ we have
 \be Q_{0,0}^{0,m}(Z,Z^\dag)=
 \frac{\binom{\lambda+1}{m+1}\binom{\lambda+1}{m}}{\lambda+1}\left(\cos^{2}\frac{\vartheta_+}{2}\cos^{2}\frac{\vartheta_-}{2}\right)^{\lambda-m}
 \left(\sin^{2}\frac{\vartheta_+}{2}\sin^{2}\frac{\vartheta_-}{2}\right)^{m},\label{Qm}\ee
 which has a binomial-like structure [similar to \eqref{binomdist}] with parameter $\cos^{2}\frac{\vartheta_+}{2}\cos^{2}\frac{\vartheta_-}{2}$.  
 The general case \eqref{husibasis} has a cumbersome explicit expression in terms of angles \eqref{paramang}, but closed expressions can still be written for particular matrices $Z$. In fact, for 
 diagonal  $Z= \Xi=\begin{pmatrix} \xi_+ &0\\ 0&  \xi_- \end{pmatrix}, \xi_\pm=\tan\frac{\vartheta_\pm}{2}e^{\ic\beta_\pm},$ and using that 
 $\cD^{j}_{q_a,q_b}(\Xi)=\delta_{q_a,q_b}\xi_+^{j+q_a}\xi_-^{j-q_a}$ we arrive at 
 \bea 
 Q_{q_a,q_b}^{j,m}(\Xi,\Xi^\dag)&=&\delta_{q_a,q_b}\frac{(2j+1)}{\lambda+1}\binom{\lambda+1}{2j+m+1}\binom{\lambda+1}{m}\nn\\ &&\times 
 \left(\cos^{2}\frac{\vartheta_+}{2}\cos^{2}\frac{\vartheta_-}{2}\right)^{\lambda-j-m}
 \left(\sin^{2}\frac{\vartheta_+}{2}\sin^{2}\frac{\vartheta_-}{2}\right)^{j+m}
 \left(\frac{\tan^2\frac{\vartheta_+}{2}}{\tan^2\frac{\vartheta_-}{2}}\right)^{q_a}.\nn\eea
 Other simplifications are also possible for the cases:
 \bea
 &\cD^{j}_{j,j}(Z)=(z^0+z^3)^{2j},\, \cD^{j}_{-j,-j}(Z)=(z^0-z^3)^{2j},&\nn\\ & \cD^{j}_{j,-j}(Z)=(z^1-\ic z^2)^{2j},\, \cD^{j}_{-j,j}(Z)=(z^1+\ic z^2)^{2j}.&
 \eea
 
This matrix generalization of the binomial structure is also apparent from a 
 Fock-space representation  of isospin-$\lambda$
CS  \eqref{u4cs} given in Ref. \cite{GrassCSBLQH}. In this Fock-space picture, boson annihilation operators $a$ and $b$ in \eqref{su2BE} are replaced by their matrix counterparts 
\be
\mathbf{a}=\begin{pmatrix} a_1^\downarrow & a_2^\downarrow \\ a_1^\uparrow & a_2^\uparrow
\end{pmatrix},\; \mathbf{b}= \begin{pmatrix} b_1^\uparrow & b_2^\uparrow\\ b_1^\downarrow & b_2^\downarrow
\end{pmatrix}, \ee
where $\ell_{1,2}^{\uparrow,\downarrow}$ annihilate flux quanta attached  to  spin $\uparrow$ and $\downarrow$ electrons 1 and 2 in layer $\ell=a,b$ in a BLQH context. 
If denote by $\check{\mathbf{a}}=\um\eta^{\mu\nu}\tr(\sigma_\mu\mathbf{a})\sigma_\nu$ and
$\check{\mathbf{b}}=\um\eta^{\mu\nu}\tr(\sigma_\mu\mathbf{b})\sigma_\nu$, then the 
CS $|Z\ra$ in  \eqref{u4cs} can be written as a boson condensate of $2\lambda$ flux quanta 
\be
|Z\ra=\frac{1}{\lambda!\sqrt{\lambda+1}}\left(\frac{\det(\check{\mathbf{b}}^\dag+
Z^\dag\check{\mathbf{a}}^\dag)}{\sqrt{\det(\sigma_0+ZZ^\dag)}}\right)^\lambda|0_F\ra
\label{u4csfock}
\ee
and the Jordan-Schwinger boson realization of the sixteen $U(4)$ isospin operators ${\tau}_{\mu\nu}$ in terms of eight boson operators $\mathbf{a}$ and $\mathbf{b}$ is written as 
${T}_{\mu\nu}=\tr({\mathcal Z}^\dag\tau_{\mu\nu} \mathcal Z)$  
with $\mathcal Z=(\mathcal Z_1,\mathcal Z_2)=\begin{pmatrix} \mathbf a\, \mathbf b\end{pmatrix}^t$ a ``two-fermion compound''. We have proved in \cite{CalixtoJPCM14} that wave functions are antisymmetric (fermionic character) under the
interchange of the two electrons $(\mathcal{Z}_1,\mathcal{Z}_2)\to (\mathcal{Z}_2,\mathcal{Z}_1)$ for $\lambda$ odd, and they are symmetric (bosonic character) for $\lambda$ even. 

\subsection{Operator symbols: BLQH order parameters}\label{symbolsec}

Now we comment on the analytic picture provided by the CS (Fock-Bargmann) representation $\Psi(Z)$  of any state $|\psi\rangle\in {\cal H}_\lambda$, defined as 
\begin{equation}
 \Psi(Z)\equiv K_{\lambda/2}(Z,Z^\dag)\langle Z|\psi\rangle.\label{FBrep2}
\end{equation}
For example, the basis states $|\psi\rangle=|{}{}_{q_a,q_b}^{j,m}\ra$ are represented by the homogeneous polynomials  $\varphi_{q_a,q_b}^{j,m}(Z)$ of degree $2j+2m$ in four complex variables 
$z^\mu$, whereas general states $|\psi\ra$ are represented by polynomials $\Psi(Z)$ of degree $\leq 2\lambda$. 
This makes the space  ${\cal H}_\lambda(\mathbb G_2)$ of polynomials $\Psi(Z)$ a {reproducing kernel Hilbert space}  with measure 
$d\mu_\lambda(Z,Z^\dag)\equiv d\mu(Z,Z^\dag)/K_{\lambda}(Z,Z^\dag)$. Inside this 
CS picture, isospin $U(4)$ operators $T_{\mu\nu}$ are represented by differential operators $\mathcal{T}_{\mu\nu}$. 
Writing $Z=z^\mu\sigma_\mu$, 
$z^2=z_\mu z^\mu$, $\partial_\mu=\partial/\partial z^\mu$ and
$\check\partial_\mu=\partial/\partial z_\mu=\partial^\mu$,
these differential operators have the following expression [we use the notation commented before \eqref{spinab}]
\be
\mathcal{M}_{\mu\nu}=z_\mu \partial_\nu-z_\nu \partial_\mu,\;
\mathcal{T}_{30}=2(z^\mu\partial_\mu-\lambda),\;
\mathcal{T}_{-\mu}=\check\partial_\mu, \; 
\mathcal{T}_{+\mu}= z^2\check{\mathcal{T}}_{-\mu}-z_\mu \mathcal{T}_{30},\label{difop}
\ee
so that $\mathcal{T}_{\mu\nu}\Psi(Z)= K_\lambda(Z,Z^\dag)\langle Z|{T}_{\mu\nu}|\psi\rangle$. As already commented in section \ref{Bloch}, this analytic representation 
reduces the computation of (technically elaborate in principle) matrix elements like $\la Z|T_{\mu\nu}|Z'\ra$ to simple derivatives 
\begin{equation}
 \la Z|T_{\mu\nu}|Z'\ra=\frac{\mathcal{T}_{\mu\nu}K_{\lambda}(Z,Z'^\dag)}{K_{\lambda}(Z,Z^\dag)K_{\lambda}(Z',Z'^\dag)},
\end{equation}
with $K_{\lambda}(Z,Z'^\dag)=(1+z^\mu\delta_{\mu\nu}\bar z'^\nu+z^\mu\eta_{\mu\nu}\bar z'^\nu)^\lambda$. Here we want to analyze the operator symbols 
$\la T_{\mu\nu}\ra\equiv \la Z|T_{\mu\nu}|Z\ra=K_{\lambda}^{-1}(Z,Z^\dag)\mathcal{T}_{\mu\nu}K_{\lambda}(Z,Z^\dag)$, which turn out to be
\bea
&\la{T}_{30}\ra= 2\lambda\frac{-1+\det(Z^\dag Z)}{\det(\sigma^0+Z^\dag Z)},\; \la {M}_{\mu\nu}\ra=2\lambda\frac{
z_\mu \bar z^\nu-z_\nu \bar z^\mu}{\det(\sigma^0+Z^\dag Z)}, & \nn \\ &\la{T}_{-\mu}\ra= 2\lambda\frac{\bar z_\mu+\det(Z^\dag)\check z_\mu}{\det(\sigma^0+Z^\dag Z)},\; 
\la {T}_{+\mu}\ra= 2\lambda\frac{ z_\mu+\det(Z)\bar z^\mu}{\det(\sigma^0+Z^\dag Z)}.& \label{T3MTpm}
\eea
We can check that $\la {T}_{+\mu}\ra=
\det(Z)\la\check{{T}}_{-\mu}\ra-z_\mu\la{T}_{30}\ra$ and  
$\la {M}_{\mu\nu}\ra=z_\mu \la\check{{T}}_{-\nu}\ra-z_\nu \la\check{{T}}_{-\mu}\ra$. In the same way, we 
can also compute symbols of quadratic (and more general) operator combinations 
$\la T_{\mu\nu}T_{\mu'\nu'}\ra=K_{\lambda}^{-1}(Z,Z^\dag)\mathcal{T}_{\mu\nu}(\mathcal{T}_{\mu'\nu'}K_{\lambda}(Z,Z^\dag))$. For example, 
$U(2)^2$-invariant (i.e., commuting with $M_{\mu\nu}$) quadratic operators: $T_{30}^2$, $M^2$ and 
\bea
\um(T_{+\mu}T_{+}^\mu+T_{-\mu}T_{-}^\mu)&=& 4(P_1^2-P_2^2-\vec{R}_1^2+\vec{R}_2^2)\nn\\
\frac{1}{2\ic}(T_{+\mu}T_{+}^\mu-T_{-\mu}T_{-}^\mu)&=& 4(P_1P_2-\vec{R}_1\cdot\vec{R}_2)\\
\check{T}_{-\mu}T_+^\mu&=& 
\vec{P}^2-P_3^2+\mathbf{R}^2-\vec{R}_3^2-4P_3,\nn
\eea
have the following symbols
\bea
 & \la{T}_{30}^2\ra=\frac{\lambda-1}{\lambda}\la{T}_{30}\ra^2+
4\lambda\frac{1+\det(Z^\dag Z)}{\det(\sigma^0+Z^\dag Z)},
\; \la{T}_{-\mu}{T}_-^\mu\ra=\frac{4\lambda(\lambda+1)\det(Z^\dag)}{\det(\sigma^0+Z^\dag Z)},\;
\la{T}_{+\mu}{T}_+^\mu\ra=\frac{4\lambda(\lambda+1)\det(Z)}{\det(\sigma^0+Z^\dag Z)},&\nn\\ 
 &\la\check{{T}}_{-\mu}{T}_+^\mu\ra=2\left(\lambda+
\lambda \frac{\lambda+2+(\lambda-2)\det(Z^\dag Z)}{\det(\sigma^0+Z^\dag Z)}-
\frac{\lambda-1}{4\lambda}\la{T}_{30}\ra^2\right),&\label{quadsymb}\\ 
&\la{M}_{\mu\nu} {M}^{\mu\nu}\ra=\frac{2\lambda+1}{2}\la{T}_{30}^2\ra +\frac{1-\lambda^2}{\lambda}
\la{T}_{30}\ra^2-2\lambda(\lambda+2)&\nn\\ &=\frac{1}{2}\la{T}_{30}^2\ra+4\la{T}_{30}\ra+
2 \la\check{{T}}_{-\mu}{T}_+^\mu\ra -2\lambda(\lambda+4),\nn
& 
\eea
where the last equality is a consequence of the quadratic Casimir \eqref{Casimir} and the fact that 
$\la{{T}}_{+\mu}\check{T}_-^\mu\ra= \la\check{{T}}_{-\mu}{T}_+^\mu\ra+
4\la{T}_{30}\ra$, which comes from the commutation relation 
 $\lb T_{+\mu},\check{T}_{-\nu}\rb = \eta_{\mu\nu} T_{30}+2M_{\mu\nu}$ and the antisymmetry of $M_{\mu\nu}$. 
 Note that all quadratic symbols \eqref{quadsymb} only depend on $U(2)^2$ invariant quantities like 
 $\tr(ZZ^\dag)=|\xi_+|^2+|\xi_-|^2$ and $\det(ZZ^\dag)=|\xi_+|^2|\xi_-|^2$. In \cite{CalixtoJPCM14} we have studied the interlayer imbalance $\la P_3\ra=\la T_{30}\ra/2$ and its 
 fluctuations $\la P_3^2\ra-\la P_3\ra^2$ in a CS. Here we want to analyze general pspin $\la \vec{P}\ra$ and spin $\la \vec{S}\ra$ symbols. 
 Before, we should stress that non-zero fluctuations $\sigma_A^2=\la A^2\ra-\la A\ra^2$ of spin ($A=S$) and pspin ($A=P$)
 exist in a CS, so that $\la A^2\ra\not=\la A\ra^2$ in general. For example, it is interesting to compare the Casimir \eqref{Casimir} symbol
 \begin{equation}
  \la \vec{S}^2\ra+\la \vec{P}^2\ra+\la \mathbf{R}^2\ra=\lambda(\lambda+4),\label{casimir1}
 \end{equation}
 with the quantity
 \begin{equation}
  \la \vec{S}\ra^2+\la \vec{P}\ra^2+\la \mathbf{R}\ra^2=\lambda^2,\label{casimir2}
 \end{equation}
 so that a non-zero ``variance sum'', written as:
 \begin{equation}
 \sigma_{\vec{S}}^2+\sigma_{\vec{P}}^2+\sigma_{\mathbf{R}}^2=\lambda(\lambda+4)-\lambda^2=4\lambda,
 \end{equation}
 exists and is proportional to $\lambda$. Here we shall restrict ourselves to the study of $\la\vec{S}\ra^2$ and $\la\vec{P}\ra^2$, leaving the analysis of fluctuations 
 aside.

 \begin{table} \begin{center}
 \begin{tabular}{|c|c|c|c|c|}
  \hline
 Phase:& Spin & Pspin & Canted\\ 
\hline\
Order & $\la\vec{S}\ra^2=\lambda^2$ &$\la\vec{S}\ra^2=0$ &$\la\vec{S}\ra^2\not=0$\\
parameter:& $\la\vec{P}\ra^2=0$ &$\la\vec{P}\ra^2=\lambda^2$ &$\la\vec{P}\ra^2\not=0$\\
\hline
 \end{tabular}
 \end{center}
 \caption{\label{tabla} Spin and pspin coherent mean values in the three BLQH phases.}
\end{table}
The ground state of a BLQH system at one Landau site and at filling factor $\nu=2$ is a coherent state $|Z\rangle$. Depending on the strength 
of interlayer tunneling, bias voltage, Zeeman, Coulomb, etc, interactions, the BLQH system can appear in three phases:  Spin or ferromagnetic  (pspin singlet) phase, 
 pspin (spin singlet) phase and  canted or antiferromagnetic phase \cite{EzawaBook}. The three phases are characterized by the order parameters $\la\vec{S}\ra^2$ and 
$\la\vec{P}\ra^2$ with values in Table \ref{tabla}. In the literature (see e.g. \cite{EzawaBook}), only the case $\lambda=1$ is considered for this phase diagram  
study. Here we shall determine the regions of $\mathbb{G}_2$ (i.e., the values of $Z$) associated to each of the three phases:
\begin{enumerate}
 \item {\it Spin phase}:  $\la\vec{S}\ra^2=\lambda^2$. From eq. \eqref{casimir2}, we see that this condition is equivalent to $\la\vec{P}\ra^2=0=\la\mathbf{R}\ra^2$. 
 Remember that $\la {S}_i\ra=\ic\epsilon_{ikl}\la {M}_{kl}\ra$. Using the mean values $\la M_{\mu\nu}\ra$ 
 in eq. \eqref{T3MTpm} and the parametrization \eqref{paramang}, we can write $\la\vec{S}\ra^2=\lambda^2 f(\vartheta_+,\vartheta_-)g(\theta_a,\theta_b,\Delta\phi)$, with 
 \begin{eqnarray}
 f(\vartheta_+,\vartheta_-)&=& \frac{1}{4}(\cos\vartheta_--\cos\vartheta_+)^2,\label{fg}\\
 g(\theta_a,\theta_b,\Delta\phi)&=&\um (1-\cos\theta_a\cos\theta_b-\cos\Delta\phi\sin\theta_a\sin\theta_b),\nn
 \end{eqnarray}
and $\Delta\phi=\phi_a-\phi_b$. The range of $f$ and $g$ is the interval $[0,1]$. The condition $\la\vec{S}\ra^2=\lambda^2$ then implies 
$f(\vartheta_+,\vartheta_-)=1=g(\theta_a,\theta_b,\Delta\phi)$. The condition $f(\vartheta_+,\vartheta_-)=1$ is fulfilled 
for $(\vartheta_-,\vartheta_+)=(\pi,0)$ or $(0,\pi)$.  The condition $g(\theta_a,\theta_b,\Delta\phi)=1$ is fulfilled for example 
for $\theta_a+\theta_b=\pi$ and $\Delta\phi=\pi$, although some other possibilities exist.  

\item {\it Pspin phase}:  $\la\vec{P}\ra^2=\lambda^2$. Remember that  $\la {P}_3\ra=\frac{\la T_{30}\ra}{2}$ and 
 \begin{eqnarray}
 \la {P}_1\ra&=&\frac{\la {T}_{+0}\ra+\la {T}_{-0}\ra}{2}={\lambda}\frac{\Re[\tr(Z)(1+\det(Z^\dag)]}{\det(\sigma_0+ZZ^\dag)},\nn\\ 
 \la {P}_2\ra&=&\frac{\la {T}_{+0}\ra-\la {T}_{-0}\ra}{2\ic}={\lambda}\frac{\Im[\tr(Z)(1-\det(Z^\dag)]}{\det(\sigma_0+ZZ^\dag)}.
\end{eqnarray}
 From eq. \eqref{casimir2}, we see that this condition is equivalent to $\la\vec{S}\ra^2=0=\la\mathbf{R}\ra^2$. In particular, from \eqref{fg}, we see that 
 $\la\vec{S}\ra^2=0$ if $f=0$ or $g=0$. We have that $f=0$ if $\vartheta_+=\vartheta_-=\vartheta$ and that $g=0$ when, for example, $\theta_a=\theta_b$ and $\Delta\phi=0$ (other 
 solutions are possible). In these particular cases we have that $\la\vec{P}\ra^2=\lambda^2(1-\sin^2[(\beta_--\beta_+)/2]\sin^2\vartheta)$, which fulfills 
 $\la\vec{P}\ra^2=\lambda^2$ when $\vartheta=0$ (that is, $Z=0$) or when $\beta_-=\beta_+$ (i.e., $Z=\tan(\vartheta/2)\sigma_0$).
  \item {\it Canted phase}:  $\la\vec{S}\ra^2\not=0\not=\la\vec{P}\ra^2$. This condition is accomplished almost everywhere in $\mathbb{G}_2$, except for the zero-measure 
  sets defining spin and pspin phases.
\end{enumerate}

A deeper study of BLQH Hamiltonians and their phase diagrams will be done elsewhere.

\subsection{Husimi function and Wehrl's entropy: a conjecture}\label{Husisec}

The Husimi function of a given state $|\psi\ra\in \mathcal{H}_\lambda$ is
\be
Q_\psi(Z,Z^\dag)\equiv|\la Z|\psi\ra|^2=\frac{|\Psi(Z)|^2}{\det(\sigma_0+ZZ^\dag)^{\lambda}}.
\ee
For unit norm states $\la\psi|\psi\ra=1$, $Q_\psi$ is normalized according to $\int_{\mathbb{G}_2}Q_\psi(Z,Z^\dag)d\mu(Z,Z^\dag)=1$, with measure \eqref{measureG2}. 
At the end of Section \ref{Bloch} we have commented on Wehrl's and Lieb's conjectures on the entropy of ordinary and spin-$s$ CS, respectively. 
In Ref. \cite{Wehrl79}, Wehrl conjectured that any Glauber (harmonic oscillator) coherent
state has a minimum Wehrl entropy of $W_\mathrm{min}=1$. This conjecture was proved by Lieb \cite{LiebCMP}, who also conjectured that the extension
of Wehrl's definition of entropy to spin-$s$ CS $|z\ra$ will yield a minimum entropy of $W_\mathrm{min}=1-1/(2 s + 1)$ 
(note that $W_\mathrm{min}\to 1$ for high spin values). This conjecture has been 
recently proved by Lieb and Solovej in \cite{LiebAM}, and an extension to $SU(N)$ coherent states seems to be also plausible, at least for the  
fully symmetric representation (complex-projective $\mathbb{C}P^{N-1}$ CSs).

Here we extend the Wehrl's definition of entropy of a state $\psi$ to Grasmannian isospin-$\lambda$ CSs $|Z\ra$ as
\be
W_\psi=-\int_{\mathbb{G}_2}Q_\psi(Z,Z^\dag)\ln Q_\psi(Z,Z^\dag) \,d\mu(Z,Z^\dag).\ee
This quantity is a measure of the area occupied by $\psi$ in phase space. We conjecture that this area is minimum when $\psi$ is a CS 
(i.e.,  $|\psi\ra=|Z'\ra$ for any $Z'\in\mathbb{G}_2$) and the minimum value is given by
 \be
 W_{\mathrm{min}}=\frac{4\lambda(2+\frac{\sqrt{2}}{2}+\lambda)(2-\frac{\sqrt{2}}{2}+\lambda)}{4(1+\lambda)(2+\lambda)(3+\lambda)}=
 4-\frac{1}{1+\lambda}-\frac{4}{2+\lambda}-\frac{3}{3+\lambda}.\label{Wmin}
 \ee
It is direct to prove that this minimum value is attained in particular by the lowest-weight state $|\psi\ra=|{}{}_{0,0}^{0,0}\ra$ (i.e., $Z'=0$), whose Husimi function is given by the 
simple expression \eqref{Qlowest}, and by the highest-weight state $|\psi\ra=|{}{}_{0,0}^{0,\lambda}\ra$ (i.e., $Z'=\infty$),  whose Husimi function is given by  
\eqref{Qm} for $m=\lambda$. For other particular CSs $|Z'\ra$, we have numerically checked that they also attain the minimum Wehrl's entropy \eqref{Wmin}. A formal proof 
of this conjecture will be investigated elsewhere. 

Remember that, for $U(2)/U(1)^2$-CSs we had $W_{\mathrm{min}}\to 1$ for high spin $s$ values. Now, for $U(4)/U(2)^2$-CSs we have $W_{\mathrm{min}}\to 4$ for high isospin $\lambda$ values. 
This limiting entropic value coincides with the complex dimension $d_\mathbb{C}(\mathbb{P})$ of the corresponding phase space $\mathbb{P}$, namely: 
$d_\mathbb{C}({\mathbb{S}^2})=1$ for the 
two-sphere  and $d_\mathbb{C}({\mathbb{G}_2})=4$ for the Grassmannian.

Another measure of the area occupied by $\psi$ in phase space is the Husimi second moment:
\be
M_\psi=\int_{\mathbb{G}_2}Q_\psi^2(Z,Z^\dag)d\mu(Z,Z^\dag),\label{MQ}\ee
which has to do with the so called ``inverse participation ratio'' (IPR). Broadly speaking, the IPR measures the 
spread of a state $|\psi\rangle$ over a basis  $\{|i\rangle\}_{i=1}^d$. More precisely, if $p_i$ is the 
probability of finding the (normalized) state $|\psi\rangle$ in $|i\rangle$, then the IPR is defined as $M_\psi=\sum_i p_i^2$. 
If $|\psi\rangle$ only ``participates'' of a single state $|i_0\rangle$, then 
$p_{i_0}=1$ and $M_\psi=1$ (large IPR), whereas if $|\psi\rangle$ equally participates on all of them (equally distributed), $p_{i}=1/{d}, \forall i$, then 
$M_\psi=1/d$ (small IPR). Therefore, the IPR is a measure of the localization of $|\psi\rangle$ in the corresponding basis. For our case, the Husimi second moment 
\eqref{MQ} measures how close is $\psi$ to a coherent state $|Z'\ra$. We also conjecture that the maximum Husimi second moment 
\be
 M_{\mathrm{max}}=\frac{(2+\lambda)^2(3+\lambda)}{4(1+\lambda)(1+2\lambda)(3+2\lambda)}=\frac{1}{16}-\frac{1/2}{1+\lambda}
 +\frac{45/32}{1+2\lambda}+\frac{3/32}{3+2\lambda}
 \ee
is attained when $\psi$ is a CS. Remember that, for $U(2)/U(1)^2$-CSs we had $M_{\mathrm{max}}\to 1/2$ for high spin $s$ values. Now, for $U(4)/U(2)^2$-CSs we have 
$M_{\mathrm{max}}\to 1/16$ for high isospin $\lambda$ values. 
This limiting IPR value coincides with $2^{-D}$, with $D=d_\mathbb{C}(\mathbb{P})$  the complex dimension of the corresponding phase space $\mathbb{P}$.

\section{Conclusions and outlook}\label{comments}

We have analyzed the structure and properties of a previously introduced family of CS labeled by points $Z$  ($2\times 2$ complex matrices) in the Grassmannian $\mathbb{G}_2=U(4)/U(2)^2$.  
Applications to bilayer quantum Hall systems at fractions of filling factor $\nu=2$ have been commented all along the paper in order to exemplify some abstract mathematical constructions. 
It is interesting to compare the Grassmannian CS in eqs. \eqref{u4cs} and \eqref{u4csfock} with the Bloch CS  in eqs. \eqref{su2cs}
and \eqref{su2BE}, together with the CS overlaps \eqref{u4csov} and \eqref{su2overlap}, respectively. We perceive a similar structure between
them, although the Grassmannian  case is more involved and constitutes a kind of ``matrix $Z$ generalization of the scalar $z$''. The matrix noncompact $U(2,2)$ version of $U(1,1)$ CS has also been explored 
by us in  \cite{EMSMTA,OscRealConfPart}, and applications to massive conformal particles and pairing systems have been commented. 

We have also extended Wehrl's definition of entropy 
to Grassmannian CS and conjectured a lower bound for it (resp. an upper bound for the Husimi second moment). The study of localization properties of Hamiltonian eigenfunctions is also an important 
subject and we believe that our result will be relevant here. Actually, CS in general are essential to study the classical, thermodynamic 
of mean-field limit of algebraic quantum models undergoing a quantum phase transition, providing  order parameters and the corresponding phase diagram. As already stated at the introduction, 
the Husimi function and its entropic measures have proved to be useful in the characterization of quantum phase transitions of  
several molecular and optical models \cite{DickeHusimiPRA,VibronHusimiPRA,LipkinPS,LipkinEPL,DickePS} and topological insulators \cite{calixto15}. 
The use of Grassmannian CS as variational states to approximate the ground state energy of bilayer quantum Hall systems is in progress. We hope that they will provide new 
insight into the subject.

\section*{Acknowledgements}

 The work was supported by 
the  Spanish Projects:  MINECO FIS2014-59386-P and the Junta de Andaluc\'{\i}a projects P12.FQM.1861 and FQM-381.

\end{document}